\documentstyle[12pt]{article}
\begin{document}
\baselineskip=7mm
%%%%%%%%%%%%%%%%%%%%%%%%%%%%

\noindent
SAGA-HE-102-96, May 1996

\bigskip

\bigskip

\noindent
{\bf Properties of nuclear matter in cut-off field theory and renormalization group methods}

\bigskip

\centerline{By}

\noindent
{\bf H. Kouno, T. Mitsumori, Y. Iwasaki, K. Sakamoto, N. Noda, }

{\bf K. Koide and A. Hasegawa}

\noindent
{Department of Physics, Saga University, Saga 840, Japan}

\centerline{and}

\noindent
{\bf M. Nakano}

\noindent
{University of Occupational and Environmental Health, Kitakyushu 807, Japan}

\bigskip

\centerline{\bf PACS numbers: 21.65.+f}

\bigskip

\bigskip

\centerline{\bf Abstract}

\noindent
The properties of nuclear matter are studied in the cut-off field theory. 
It is found that, under the Hartree approximation, the small cut-off makes the equations of state hard, especially at higher densities. 
The theory is modified in the framework of the renormalization group methods with arbitrary cut-off $\Lambda^\prime$. 
It is found that the expansion in terms of the $\sigma$ meson field is more favorable than the naive expansion of the inverse of $\Lambda^\prime$, when we do not use very large $\Lambda^\prime$. 

\vfill\eject

%%%%%%%%%%%%%%%%%%%%%%%%%%%%%%%%%%%%%%%%%%%%%%%%%%%%%%%%%%%%%%%%%%%%%%%%%%%

In recent two decades, nuclear matter has been studied in the framework of quantum hadrodynamics (QHD). The meson mean-field theory for nuclear matter [1] has made successful results to account for the saturation properties at the normal nuclear density. 
Following to those successes, many studies and modifications are done in the relativistic nuclear models. 
One of those modifications is inclusion of vacuum fluctuation effects, which cause divergences of physical quantities as thye are naively calculated. 
Chin [2] estimated the vacuum fluctuation effects in the Hartree approximation by using the renormalization procedures, and found that the vacuum fluctuation effects makes the incompressibility of nuclear matter smaller and closer to the empirical value than in the original Walecka model. 
The renormalization procedures are used in the other studies (e.g., see [3][4] ). 
However, it becomes more difficult to do the renormalization as the model becomes more complicated, since the renormalization procedures need analytical studies to some extent. 

On the other hand, the relation between QHD and the underlying fundamental theory, i.e., QCD, is the open question. 
One may wonder whether QHD is valid in very high-energy scale or not. 
If QHD is valid only under some energy scale, it is natural to introduce the cut-off or the form factor (e.g., see [3]) into the theory. 
Cohen [5] introduced the four dimensional cut-off into the relativistic Hartree calculation and found that the vacuum energy contribution is somewhat different from the one in the renormalization procedures, if the cut-off is not so large. 
In this paper, we studies the nuclear matter properties using the cut-off field theory of Cohen. 
After that we modifies the theory in the framework of the renormalization group methods [6][7][8] to prepare for calculations in more complicated and realistic models. 

We start with the following renormalizable Lagrangian of $\sigma$-$\omega$ model together with a regulator that truncates the theory's state space at some large $\Lambda$. 
%%%%%%%%
$$
L={\bar{\psi}}(i\gamma_\mu\partial^\mu-M+g_s\phi-g_v\gamma_\mu V^\mu )\psi
+{{1}\over{2}}\partial_\mu\phi\partial^{\mu}\phi-{{1}\over{2}}m_s\phi^2
-{{1}\over{4}}F_{\mu\nu}F^{\mu\nu}+{{1}\over{2}}m_v^2V_\mu V^\mu
-U(\phi );$$
$$ U(\phi )=\sum_{n=0}^{4}{{C_n}\over{n!}}(g_s\phi )^n, \eqno{(1)} $$
%%%%%%%%
where $\psi$, $\phi$, $V_\mu$, $M$, $m_s$, $m_v$, $g_s$, and $g_v$ are nucleon field, $\sigma$-meson field, $\omega$-meson field, nucleon mass, $\sigma$-meson mass, $\omega$-meson mass, $\sigma$-nucleon coupling, and $\omega$-nucleon coupling, respectively. 
$C_n$ is a constant parameter which is adjusted to reproduce the physical conditions as explained below. 
The Lagrangian (1) is valid only in the region of the energy scale which is smaller than $\Lambda$. 

Next we formulate the cut-off field theory according to Cohen [5]. 
In the relativistic Hartree approximation with the cut-off $\Lambda$, the one-loop contribution to the $\sigma$ effective potential is given by
%%%%%%%%
$$ U_{1-loop}(M^*,\Lambda )=i\int {{dk^4}\over{(2\pi)^4}}{{1}\over{2}}{\rm Tr}\big( \log{\big( {{k_E^2+M^{*2}}\over{\mu^2}} \big) }\big) \Theta (\Lambda^2-k_E^2), \eqno{(2)} $$
%%%%%%%%
where $M^*=M-\Phi=M-g_s<\phi >$, $\mu$ is an arbitrary scale parameter with dimensions of mass, $\Theta$ is the step function, and the subscript $E$ denotes that the momentum with it is written in the Euclidian notation. 
We choose parameter $C_n$'s to reproduce the following conditions as in the ordinary renormalization procedures [2][9]. 
%%%%%%%%
$$ {{d^n}\over{d\Phi ^n}}[U(\Phi )+U_{1-loop}(\Phi )]\vert_{\Phi=0}=0.~~~~(n=0,1,2,3,4) \eqno{(3)} $$
%%%%%%%%
Note that the different conditions give the different physical results as is pointed out by Heide and Rudaz [10]. 

The condition gives 
%%%%%%%%
$$  C_n=-U_{1-loop}^{n}(0)=-{{d^n}\over{d\Phi^n}}U(\Phi)_{1-loop}\vert_{\Phi =0},~~~~~(n=0,1,2,3,4). \eqno{(4)} $$ 
%%%%%%%%
The condition (4) with $n=0$ ensure that energy density $\epsilon$ of the system becomes zero at zero baryon density ($\rho =0$ ) and remove the $\mu$ dependence of $\epsilon$. 
The condition (4) with $n=1$ ensure that the scalar density $\rho_s$ of the nucleons becomes to zero at $\rho=0$ and $M^*=M$ at $\rho=0$. 
The conditions (4) with $n=2$ ensure that the physical mass of $\sigma$-meosn is $m_s$. 
The conditions (4) with $n=3$ and $n=4$ means that the effective cubic and quartic couplings vanish. 
After $C_n$ is determined, the scalar part of the nucleon self-energy is calculated by the equation of motion for $\sigma$ meson, i.e., 
%%%%%%%%
$$ {{\partial (L-U_{1-loop})}\over{\partial \Phi}}=0.   \eqno{(5)} $$
%%%%%%%%
The vector part of the nucleon self-energy is calculated by the equation of motion of $\omega$-meson as usual. 
If we consider the limit $\Lambda\rightarrow \infty$, the cut-off model is equivalent to the ordinary model in the renormalization procedures [2][9]. 

Using the coefficients (4), we have done self-consistent calculation in the Hartree approximation with the cut-off $\Lambda$. 
The coupling constants $g_s$ and $g_v$ are determined to reproduce the saturation properties of the nuclear matter. 
We use $k_{F0}=1.42$fm$^{-1}$ at the normal density $\rho_0$ and -15.75MeV as the binding energy $E_{b0}$ at $\rho_0$. 
In fig. 1(solid line), we show $\Lambda$ dependence of the vacuum contribution to the binding energy at the normal baryon density. 

%%%%%%%%%%%%%%%%%%%%%%%%%%%%%%%%

\centerline{$\underline{~~~~~~~}$}

\centerline{Fig. 1}

\centerline{$\underline{~~~~~~~}$}

%%%%%%%%%%%%%%%%%%%%%%%%%%%%%%%%

As is already pointed out by Cohen [5], the vacuum contribution to the energy becomes negative for the small $\Lambda$ region. 
In the region of $\Lambda\approx 0.4\sim 1.9$GeV, the absolute value of the negative vacuum energy is so large that we could not find the parameters $g_s$ and $g_v$ for the saturation conditions. 
In that region, we draw the dotted line as the guide for eyes. 
In the very small region of $\Lambda(<0.4$GeV), the vacuum energy show the complicated behavior. 
It reaches the mean field limit at $\Lambda$=0. 
However, the region may not be physical, since it is natural that $\Lambda >M^* $. 
Below we call the region of $\Lambda^>_\sim 2$GeV "available region", conveniently, and restrict our discussions to the region. 
The vacuum energy contribution is negative for the small $\Lambda$ in the available region. 
In fig. 2 (solid line), we show the $\Lambda$ dependence of the effective nucleon mass $M^*$ at the normal baryon density. 
As is already pointed out by Cohen, with large $\Lambda (>5$GeV), the value of $M^*$ is close to the result in the ordinary renormalization procedures ($\Lambda =\infty$), $0.718M$. 
In the available region ($\Lambda^>_\sim 2$GeV), the $M^*$ become smaller as $\Lambda$ becomes smaller. 
It is because the negative vacuum energy makes the coupling constant $g_v$ larger to make the repulsive effect of $\omega$-meson larger and to reproduce the saturation properties. 
The larger $g_v$ makes $M^*$ smaller, because of the Hugenholtz-van Hove theorem [11] at the normal density $\rho_0$, i.e., [12]
%%%%%%%%
$$ M^*=\sqrt{[M+E_{b0}-C_v^2\rho_0 /M^2 ]^2-k_{F0}^2}, \eqno{(6)} $$
%%%%%%%%
where $C_v=g_vM/m_v$.

%%%%%%%%%%%%%%%%%%%%%%%%%%%%%%%%

\centerline{$\underline{~~~~~~~}$}

\centerline{Fig. 2}

\centerline{$\underline{~~~~~~~}$}

%%%%%%%%%%%%%%%%%%%%%%%%%%%%%%%%

In figs. 3 and 4 (solid lines), we show the $\Lambda$ dependence of the incompressibility $K$ and skewness coefficient $K^\prime$, which are defined by 
%%%%%%%%
$$    K=9\rho_0^2{{d^2E_b}\over{d\rho^2}}\vert_{\rho =\rho_0},~~~~~K'=3\rho_0^3{{d^3E_b}\over{d\rho ^3}}\vert_{\rho =\rho_0}, \eqno{(
7)} $$
%%%%%%%%
where $E_b=\epsilon /\rho -M$. 
In our definition, large $K^\prime$ means that the equations of state becomes hard at higher densities. 
In figs. 3 and 4(solid lines), it is seen that both $K$ and $K'$ become larger as $\Lambda$ becomes smaller in the available region ($\Lambda^>_\sim 2$GeV). 
At $\Lambda \sim $2GeV, $M^*/M$ becomes very small ($\sim 0.5$) and it makes equations of state very hard. 
Especially, $K'$ becomes much larger in that region than in the large $\Lambda (> 5$GeV) region. 
This indicate that the cut-off effects may be very important at higher densities. 

%%%%%%%%%%%%%%%%%%%%%%%%%%%%%%%%

\centerline{$\underline{~~~~~~~}$}

\centerline{Figs. 3, 4}

\centerline{$\underline{~~~~~~~}$}

%%%%%%%%%%%%%%%%%%%%%%%%%%%%%%%%

Next we modify the model in the framework of the renormalization group methods [6][7][8]. 
In the renormalization group methods, we introduce an arbitrary new cut-off $\Lambda^{\prime}$, which is smaller than the original cut-off $\Lambda$. 
It should be remarked that, 
differently from the original $\Lambda$, 
$\Lambda^\prime$ is introduced conveniently 
 and does not mean the limiting energy scale of the theory. 
According to Lepage [8], we construct a low energy effective Lagrangian, requiring the physical results do not depend on $\Lambda^\prime$. 
We estimate the contributions which is discarded by introducing the new cut-off $\Lambda^\prime$. 
In the case of the vacuum energy potential, it is given by
%%%%%%%%
$$ \Delta U =-\Delta L \equiv U_{1-loop}(M^*,\Lambda^2\geq k^2_E \geq \Lambda^{\prime 2})
$$
$$
=i\int {{dk^4}\over{(2\pi)^4}}{{1}\over{2}}{\rm Tr}[ \log{\big( {{k_E^2+M^{*2}}\over{\mu^2}} \big) }] [\Theta (\Lambda^2-k_E^2)-\Theta (\Lambda^{\prime 2}-k_E^2)]
$$
$$
=-{{g}\over{8\pi^2}}\int_{\Lambda^\prime}^{\Lambda}dk_Ek_E^3\log{\big( {{k_E^2+M^{*2}}\over{\mu^2}}\big)}
=-{{g\Lambda^{\prime4}}\over{8\pi^2}}\int_1^{\Lambda/\Lambda^\prime}dx\log{\big( {{x^2+y^2}\over{(\mu/\Lambda^\prime )^2}}\big)}, \eqno{(8)} $$
%%%%%%%%
where $x=k_E/\Lambda^\prime$, $y=M^*/\Lambda^\prime$, and $g$ is the degeneracy factor ($g=4$ in the nuclear matter). 
If $M^*/\Lambda^\prime <1$, we could expand the last line of eq. (8) around $x^2$. 
%%%%%%%%
$$ \Delta U=\sum_{m=0}^{\infty}{{1}\over{m!}}{{\partial^m( \Delta U)}\over{\partial (y^{2})^m}}\vert_{y^2=0}(y^2)^{m}
=\sum_{m=0}^{\infty}{{1}\over{m!}}{{\partial^m (\Delta U)}\over{\partial (y^2)^{m}}}\vert_{y^2=0}\big( {{M-\Phi}\over{\Lambda^\prime}}\big) ^{2m}. \eqno{(9)} $$
%%%%%%%%
Therefore, if we use the new cut-off $\Lambda^{\prime}$, we must add $\Delta L$ to the Lagrangian as the effective potential, to keep that physical results do not depend on $\Lambda^\prime$. 
$L+\Delta L$ is the low energy effective Lagrangian under the energy scale $\Lambda^\prime$. 
Since the coefficient of the expansion (9) is order $(\Lambda^\prime )^4$, the $m$-th term of (9) has the $(1/\Lambda^\prime )^{2m-4}$ order contribution if we treat $\Lambda^\prime$ as the same order as the $\Lambda$. 
In actual calculations, we truncate the Taylor expansion (9) at some finite maximum $m$. 
In that case, the calculations include $O((1/\Lambda^\prime )^{2m-2})$ errors. 
To get higher accuracy, higher order terms in the Taylor expansion (9) are needed. 

What is the advantage of using the low energy effective Lagrangian $L+\Delta L$ with $\Lambda^\prime$ instead of the original one with the original cut-off $\Lambda$? 
In the Hartree approximations, using $L+\Delta L$ has little advantage, since the momentum dependence of the self-energies or the vertex functions are not calculated. 
However, in the modified model such as the Hartree-Fock approximation, it is very useful to introduce new cut-off and consistently neglect the momentum dependence of those quantities in high momentum region $(k_E^2 > \Lambda^{\prime 2})$. 
Therefore, it is important to study the efficiency of the expansion (9), since the Hartree contributions are very important even in the modified model. 

In figs. 1$\sim$4(dashed lines), we show the $\Lambda^\prime$ dependence of several quantities, using the expansion (9) with truncating $m$ at 3. 
This means we do the calculations with $O((1/\Lambda^\prime )^{4})$ errors. 
In those calculations, we put $\Lambda =10$GeV, and coupling constants $g_s$ and $g_v$ are determined to reproduce the saturation properties at $\Lambda^\prime =\Lambda$. 
In those figures, we see that, for small $\Lambda^\prime (< 5$GeV), the result somewhat deviates from the exact result at $\Lambda^\prime =\Lambda$, and the divergent behavior is seen in the smaller $\Lambda^\prime (< 2$GeV) region, where we do not show the results. 
It is because each term of the expansion (9) has the divergent behavior in the limit $\Lambda^\prime \rightarrow 0$, although $\Delta U$ itself does not diverge if we do not truncate the expansion at some finite $m_{max}$. 
In figs. 1$\sim $4(dashed-dotted lines), we show the result with $O((1/\Lambda^\prime )^{6})$ errors($m_{max}=4$). 
It is seen that divergent behavior at small $\Lambda^\prime$ is little improved by introducing the higher-order correction in (9). 

To improve the divergent behavior, we re-expand eq. (9) in terms of $\Phi$. 
%%%%%%%%
$$  \Delta U=\sum_{l=0}^{\infty} {{1}\over{l!}}{{\partial^l( \Delta U)}\over{\partial \Phi^l}}\vert_{\Phi =0}\Phi^l. \eqno{(10)} $$
%%%%%%%%
Comparing (9) with (10), it is easily seen that $l$-th term in (10) is order of $(1/\Lambda^\prime )^{2j-4}$, where $j$ is the integer part of $(l+1)/2$. 
However, different from the expansion in (9), besides the contribution of the order of $(1/\Lambda^\prime )^{2j-4}$, the additional $O((1/\Lambda^\prime )^{2j-2})$ contributions are also included in each $l$-th term. 
It is easily shown that these higher order contributions remove the divergence at $\Lambda^{\prime}=0$ in each term of (10). 
The each term of the expansion (10) has correct limit at $\Lambda^\prime =0$ as well as at the limit $\Lambda^\prime =\Lambda$. 
(Note that $\Delta U$ does not become zero at $\Lambda^\prime =0$.) 
In figs. 1$\sim$4 (bald dashed lines), we show the $\Lambda^{\prime}$-dependence of several quantities by using the expansion (10) with $l_{max}=6$. 
Although the 6-th term in the expansion (10) is order of $(1/\Lambda^\prime)^2$, calculated quantities show very stable behavior. 
Expansion (10) is more favorable than the naive $1/\Lambda^\prime$ expansion (9), if we  do not use very large $\Lambda^\prime$. 

In summary, we have studied the nuclear matter with the cut-off field theory, under the Hartree approximation. 
It is found that small cut-off makes the equations of state hard, especially at high densities. 
The theory is modified in the framework of renormalization group methods. 
Modified theory may be more useful than the naive cut-off field theory, if we calculate the momentum dependence of the self-energies and the vertex functions. 
The convergence of the expansion of the inverse of the cut-off are studied. 
It is found that $\Phi$-expansion are more useful than the naive expansion. 
It is very interesting to do calculations of the Hartree-Fock and the random phase approximation in the framework of the renormalization group methods, especially in the case of the nonrenormalizable Lagrangian. 
They are now under the studies. 

%%%%%%%%
\noindent
Acknowledgement: The authors gratefully thank K. Harada, who recommend them to use the renormalization group methods in quantum hadrodynamics, for many useful suggestions and discussions. 
They also thank M. Yahiro, T. Kohmura and H. Yoneyama for useful discussions, 
 and acknowledge the computing time granted by Research Center for Nuclear Physics (RCNP). 

\vfill\eject

%%%%%%%%
\centerline{\bf References}
%%%%%%%%

\bigskip

\noindent
%%%%%%%%%%%%%%%%%%%%%%%%%%%%%%%%%%%%%%%%%%%%%%%%%%%%%%%%%%%%%%%%%%%%%%%%%%%
[1] J.D. Walecka, Ann. of Phys. {\bf 83} (1974) 491
%%%%%%%%%%%%%%%%%%%%%%%%%%%%%%%%%%%%%%%%%%%%%%%%%%%%%%%%%%%%%%%%%%%%%%%%%%%

\noindent
%%%%%%%%%%%%%%%%%%%%%%%%%%%%%%%%%%%%%%%%%%%%%%%%%%%%%%%%%%%%%%%%%%%%%%%%%%%
[2] S.A. Chin, Phys. Lett. {\bf 62B} (1976) 263: S.A. Chin, Ann. of Phys. {\bf 108} (1977) 301
%%%%%%%%%%%%%%%%%%%%%%%%%%%%%%%%%%%%%%%%%%%%%%%%%%%%%%%%%%%%%%%%%%%%%%%%%%%

\noindent
%%%%%%%%%%%%%%%%%%%%%%%%%%%%%%%%%%%%%%%%%%%%%%%%%%%%%%%%%%%%%%%%%%%%%%%%%%%
[3] R.J. Furnstahl and C.J. Horowitz, Nucl. Phys. {\bf A485} (1988) 632.%%%%%%%%%%%%%%%%%%%%%%%%%%%%%%%%%%%%%%%%%%%%%%%%%%%%%%%%%%%%%%%%%%%%%%%%%%%

\noindent
%%%%%%%%%%%%%%%%%%%%%%%%%%%%%%%%%%%%%%%%%%%%%%%%%%%%%%%%%%%%%%%%%%%%%%%%%%%
[4] K. Tanaka and W. Bentz, Nucl. Phys. {\bf A540} (1992) 383.
%%%%%%%%%%%%%%%%%%%%%%%%%%%%%%%%%%%%%%%%%%%%%%%%%%%%%%%%%%%%%%%%%%%%%%%%%%%

\noindent
%%%%%%%%%%%%%%%%%%%%%%%%%%%%%%%%%%%%%%%%%%%%%%%%%%%%%%%%%%%%%%%%%%%%%%%%%%%
[5] T.D. Cohen, Phys. Lett. {\bf B211}(1988)384. 
%%%%%%%%%%%%%%%%%%%%%%%%%%%%%%%%%%%%%%%%%%%%%%%%%%%%%%%%%%%%%%%%%%%%%%%%%%%

\noindent
%%%%%%%%%%%%%%%%%%%%%%%%%%%%%%%%%%%%%%%%%%%%%%%%%%%%%%%%%%%%%%%%%%%%%%%%%%%
[6] K.G. Wilson and J. Kogut, Phys. Rep. {\bf 12} (1974)75. 
%%%%%%%%%%%%%%%%%%%%%%%%%%%%%%%%%%%%%%%%%%%%%%%%%%%%%%%%%%%%%%%%%%%%%%%%%%%

\noindent
%%%%%%%%%%%%%%%%%%%%%%%%%%%%%%%%%%%%%%%%%%%%%%%%%%%%%%%%%%%%%%%%%%%%%%%%%%%
[7] K. G. Wilson, Rev. Mod. Phys. {\bf 55} (1983)583. 
%%%%%%%%%%%%%%%%%%%%%%%%%%%%%%%%%%%%%%%%%%%%%%%%%%%%%%%%%%%%%%%%%%%%%%%%%%%

\noindent
%%%%%%%%%%%%%%%%%%%%%%%%%%%%%%%%%%%%%%%%%%%%%%%%%%%%%%%%%%%%%%%%%%%%%%%%%%%
[8] G P. Lepage, "What is renormalization?" in "From actions to answers", Proceedings of the 1989 theoretical advanced study institute in elementary particle physics, edited by T. DeGrand and D. Toussaint, p483, (World Scientific, Singapore 1990). 
%%%%%%%%%%%%%%%%%%%%%%%%%%%%%%%%%%%%%%%%%%%%%%%%%%%%%%%%%%%%%%%%%%%%%%%%%%%

\noindent
%%%%%%%%%%%%%%%%%%%%%%%%%%%%%%%%%%%%%%%%%%%%%%%%%%%%%%%%%%%%%%%%%%%%%%%%%%%
[9] 
B.D. Serot and J.D. Walecka, $The$ $Relativistic$ $Nuclear$ $Many$-$Body$ $Problem$ in: Advances in nuclear physics, vol. {\bf 16}, edited by J.W. Negele and E. Vogt (Plenum Press, New York, 1986).
%%%%%%%%%%%%%%%%%%%%%%%%%%%%%%%%%%%%%%%%%%%%%%%%%%%%%%%%%%%%%%%%%%%%%%%%%%%

\noindent
%%%%%%%%%%%%%%%%%%%%%%%%%%%%%%%%%%%%%%%%%%%%%%%%%%%%%%%%%%%%%%%%%%%%%%%%%%%
[10] E.K. Heide and S. Rudaz, Phys. Lett. {\bf B262}(1991)375. 
%%%%%%%%%%%%%%%%%%%%%%%%%%%%%%%%%%%%%%%%%%%%%%%%%%%%%%%%%%%%%%%%%%%%%%%%%%%

\noindent
%%%%%%%%%%%%%%%%%%%%%%%%%%%%%%%%%%%%%%%%%%%%%%%%%%%%%%%%%%%%%%%%%%%%%%%%%%%
[11] N.M. Hugenholtz and L.van Hove, Physica {\bf 24} (1958)363.
%%%%%%%%%%%%%%%%%%%%%%%%%%%%%%%%%%%%%%%%%%%%%%%%%%%%%%%%%%%%%%%%%%%%%%%%%%%

\noindent
%%%%%%%%%%%%%%%%%%%%%%%%%%%%%%%%%%%%%%%%%%%%%%%%%%%%%%%%%%%%%%%%%%%%%%%%%%%
[12] J. Boguta and and A.R. Bodmer, Nucl. Phys. {\bf A292}(1977)413. 
%%%%%%%%%%%%%%%%%%%%%%%%%%%%%%%%%%%%%%%%%%%%%%%%%%%%%%%%%%%%%%%%%%%%%%%%%%%

%%%%%%%%%%%%%%%%%%%%%%%%%%%%%%%%%%%%%%%%%%%%%%%%%%%%%%%%%%%%%%%%%%%%%%%%%%%

\vfill\eject

\centerline{\bf Figure Captions}

\bigskip

\noindent
Fig. 1 The $\Lambda$ or $\Lambda^\prime$ dependence of the vacuum contributions $E_{vac}$ to the binding energy at the normal baryon density. 
The solid line is the $\Lambda$-dependence. 
The dotted line is drawn as the guide for eyes. 
The dashed and dashed-dotted lines are the $\Lambda^\prime$-dependence calculated by truncating eq. (9) at $m_{max}=3$ and 4, respectively. 
The bald dashed line is the $\Lambda^\prime$-dependence calculated by truncating eq. (10) at $l_{max}=6$. 

\bigskip

\noindent
Fig. 2 The $\Lambda$ or $\Lambda^\prime$ dependence of the effective nucleon mass $M^*$ at the normal density. 
The meaning of each line is the same as in fig. 1. 

\bigskip

\noindent
Fig. 3 The $\Lambda$ or $\Lambda^\prime$ dependence of the incompressibility $K$. 
The meaning of each line is the same as in fig. 1. 

\bigskip

\noindent
Fig. 4 The $\Lambda$ or $\Lambda^\prime$ dependence of the skewness coefficient $K^\prime$. 
The meaning of each line is the same as in fig. 1. 

\end{document}